\documentclass[useAMS,usenatbib]{mn2e}
\usepackage{epsfig}
\usepackage{natbib}

\title[Statistical properties of NEWPS catalogue]{Statistical properties of extragalactic
sources in the New Extragalactic WMAP Point Source (NEWPS) catalogue}

\author[Gonz\'alez-Nuevo et al.]{J. Gonz\'alez-Nuevo$^{1}$\thanks{E-mail:
gnuevo@sissa.it}, M. Massardi$^{1,2}$,
  F. Arg\"ueso$^{3}$, D. Herranz$^{4}$, L. Toffolatti$^{5}$,
  \newauthor J.L. Sanz$^{4}$, M. L\'opez-Caniego$^{4,6}$ and G. De Zotti$^{7,1}$
   \\
  $^{1}$ SISSA-I.S.A.S, via Beirut 4, I-34014 Trieste, Italy \\
  $^{2}$ Australia Telescope National Facility, CSIRO, PO Box 76 Epping, NSW 1710, Australia \\
  $^{3}$ Departamento de Matem\'aticas, Universidad de Oviedo,
  Avda. Calvo Sotelo s/n, 33007 Oviedo, Spain\\
  $^{4}$ Instituto de F\'\i{sica} de Cantabria (CSIC-UC), Avda. los
  Castros s/n, 39005 Santander, Spain \\
  $^{5}$ Departamento de F\'\i{sica}, Universidad de Oviedo,
  Avda. Calvo Sotelo s/n, 33007 Oviedo, Spain \\
  $^{6}$ Astrophysics Group, Cavendish Laboratory,
  J.J. Thompson Avenue, CB3 0E1, Cambridge, United Kingdom\\
  $^{7}$ INAF-Ossevatorio Astronomico di Padova, vicolo
  dell'Osservatorio 5, I-35122 Padova, Italy}

\date{Accepted 2007 November 15.  Received 2007 November 13; in original form 2007 September 10}

\begin{document}

\maketitle

\begin{abstract}
We present results on spectral index distributions, number counts, redshift distribution and other general
statistical properties of extragalactic point sources in the NEWPS$_{5\sigma}$ sample \citep{can07}. The flux
calibrations at all the WMAP channels have been reassessed both by comparison with ground based observations and
through estimates of the effective beam areas. The two methods yield consistent statistical correction factors.
A search of the NASA Extragalactic Database (NED) has yielded optical identifications for $\sim 89\%$ of sources
in the complete sub-sample of 252 sources with $S/N\geq5$ and $S\geq 1.1$ Jy at 23 GHz; 5 sources turned out to
be Galactic and were removed. The NED also yielded redshifts for $\simeq 92\%$ of the extragalactic sources at
$|b|>10^\circ$. Their distribution was compared with model predictions; the agreement is generally good but a
possible discrepancy is noted. Using the 5 GHz fluxes from the GB6 or PMN surveys, we find that $\sim 76\%$ of
the 191 extragalactic sources with $S_{23\rm GHz}> 1.3\,$Jy can be classified as flat-spectrum sources between 5
and 23 GHz. A spectral steepening is observed at higher frequencies: only 59\% of our sources are still
flat-spectrum sources between 23 and 61 GHz and the average spectral indexes steepen from
$\langle\alpha_{5}^{23}\rangle= 0.01\pm 0.03$ to $\langle\alpha_{41}^{61}\rangle= 0.37\pm 0.03$. We think,
however, that the difference may be due to a selection effect. The source number counts have a close to
Euclidean slope and are in good agreement with the predictions of the cosmological evolution model by
\citet{dz05}. The observed spectral index distributions were exploited to get model-independent extrapolations
of counts to higher frequencies. The risks of such operations are discussed and reasons of discrepancies with
other recent estimates are clarified.
\end{abstract}


\section{Introduction} \label{sec:intro}

\begin{table*}
  \centering
  \begin{tabular}{ccccc|cc}
WMAP band \& &  Average   &  Symmetrized area &  Self-calibration correction & &&Calibrators correction\\
central freq. (GHz) & effective area (sr)                    & (sr)
& average value + error & &&   average value + error    \\ \hline\hline%
K (23)  & $(2.595 \pm 0.014) \times 10^{-4} $ & $2.46 \times 10^{-4} $ & $1.050 \pm 0.006$ &&& 0.99 $\pm$ 0.06\\
Ka (33) & $(1.569 \pm 0.012) \times 10^{-4} $ & $1.44 \times 10^{-4} $ & $1.086 \pm 0.008$ &&& 1.12 $\pm$ 0.07\\
Q  (41) & $(1.016 \pm 0.007)\times 10^{-4}  $ & $8.94 \times 10^{-5} $ & $1.136 \pm 0.008$ &&& 1.15 $\pm$ 0.08\\
V  (61) & $(4.81 \pm 0.07)\times 10^{-5}    $ & $4.19 \times 10^{-5} $ & $1.15 \pm 0.02$ &&& 1.32 $\pm$ 0.09\\
W  (94) & $(3.36 \pm 0.17) \times 10^{-5}   $ & $2.07 \times 10^{-5} $ & $(1.62 \pm 0.08)  $ &&&(1.41 $\pm$ 0.10)\\%
\hline\hline%
\end{tabular}

\caption{Results of the self-calibration process (effective area/symmetrized area; see \S~\ref{sec:auto_cal})
and of the source calibration by comparison with ground based measurements  (ground based fluxes/NEWPS fluxes or
average ratio of the corrected flux to the uncorrected one; see \S~\ref{sec:sou_cal}). The correction factors
for the W band are in parenthesis because they are very uncertain (see text).}
  \label{tb:cal_fac}
\end{table*}

The statistical properties of extragalactic sources  above $\sim 10$ GHz are still largely unknown. This is due
to the great difficulty of carrying out extensive surveys at high radio frequencies, because of the very small
fields of view of ground-based radio telescopes and of the need of relatively long on-source integrations. Up to
the end of the last century, the highest radio frequency at which large-area sky surveys had been performed was
5 GHz (GB6, \citet{gre96}; PMN, \citet{gri94,gri95}; \citet{wri94,wri96}). Recently, the situation has
considerably improved thanks to the Pilot ATCA survey that covered the declination region between $-60^\circ$
and $-70^\circ$ at 18 GHz down to 100 mJy \citep{ricci04,sadler06}, to the 9C survey covering
$520\,\hbox{deg}^2$ at 15.2 GHz to a flux limit of 25 mJy but going down to 10 mJy on small areas
\citep{waldram03,taylor01}, and to data from the VSA in the declination range $-5^\circ <\delta<60^\circ$ at 33
GHz \citep{cleary05}. The on-going Australia Telescope 20 GHz (AT20G) Survey \citep{ekers}, once completed, will
cover the whole Southern hemisphere down to 50 mJy.

Moreover, the Wilkinson Microwave Anisotropy Probe (WMAP) mission has produced the first all-sky surveys of
extragalactic sources at 23, 33, 41, 61 and 94 GHz \citep{ben03b,hin07}. >From the analysis of the first three
years survey data the WMAP team have obtained a catalogue of 323 extragalactic point sources (EPS;
\citet{hin07}), substantially enlarging the first-year one that included 208 EPS detected above a flux limit of
$\sim 0.8$-1 Jy \citep{ben03b}, with an estimated completeness limit of $\sim 1.2$ Jy. The average estimated
spectral index for detected sources is $\simeq 0.0$, thus confirming that the source class which dominates the
counts at bright fluxes is essentially made of flat-spectrum QSOs and BL Lacs, with minor contributions coming
from steep- or inverted-spectrum sources \citep{tof98,dz99,tof99,ben03b}.

In order to take full advantage of all the information content of the WMAP 3-year survey data, \citet[hereafter
LC07]{can07} have used the MHW2 filter \citep{jgn06} to obtain estimates of (or upper limits on) the flux
densities at the WMAP frequencies of a complete all-sky sample of 2491 sources at $|b|>5^\circ$, brighter than
500 mJy at 5 GHz, or at 1.4 or 0.84 GHz in regions not covered by 5 GHz surveys but covered by either the NVSS
\citep{con98} or the SUMSS (Mauch et al. 2003). This work yielded $5\sigma$ detections of 368 extragalactic
sources, including 98 sources not present in the WMAP 3-year catalogue. The results were organized in a
catalogue dubbed NEWPS$_{5\sigma}$ (New Extragalactic WMAP Point Source) Catalogue.

As pointed out by LC07, the new flux estimates showed small systematic differences, increasing with frequency,
with those obtained by the WMAP team. These differences were attributed to different approximations of the beam
shapes. In fact, the beams are complex and asymmetrical. Taking also into account that sources are observed with
different beam orientations, it is clear that the effective beam areas and, therefore, the flux calibrations,
are uncertain.

To check the calibration of LC07 fluxes we have looked for ground based measurements of NEWPS$_{5\sigma}$
sources at frequencies close to the WMAP ones, finding small but appreciable systematic differences. This
prompted us to investigate in more detail the calibration problem. The adopted methods and the results are
presented in \S\,\ref{sec:calib}. In \S\,\ref{sec:red_distrib_llf} we present the optical identifications and
the redshift distribution of our sources, based on literature data. After having recalibrated the
NEWPS$_{5\sigma}$ flux densities we have investigated the spectral index distributions in different frequency
intervals (\S\,\ref{sec:pop_spec}) and used them to obtain model independent estimates of the mm-wave source
counts, beyond the flux density intervals where they are directly determined (\S\,\ref{sec:counts}). Finally, in
\S\,\ref{sec:Conclusions}, we summarize our main conclusions.

Throughout this paper we have used the Bayes Corrected version of the NEWPS Catalogue to take into account the
well-known Eddington bias \citep[see $\S$ \ref{sec:edd_bias} or LC07 for more details]{edding}\footnote{The
Bayes-corrected version of the NEWPS Catalogue can be found at http://max.ifca.unican.es/caniego/NEWPS/}. The
spectral index, $\alpha$, is defined as $\alpha\equiv -\log(S_2/S_1)/\log(\nu_2/\nu_1)$, where $S_1$ and $S_2$
are the fluxes at the frequencies $\nu_1$ and $\nu_2$, respectively (i.e., $S_\nu \propto \nu^{-\alpha}$).

\section{Statistical flux calibration} \label{sec:calib}

We have studied in detail the calibration of the estimated NEWPS fluxes, after the application of the Bayesian
correction (see next sub-section), by adopting two independent approaches: a) comparison with ground-based
measurements at the closest frequencies; b) direct estimates of the effective beam areas for the brightest
sources.

\subsection{The Eddington bias and the Bayesian correction}\label{sec:edd_bias}

It has long been realized \citep{edding} that fluxes of sources detected above a certain $S/N$ threshold in
noisy fields are, on average, overestimated since there are similar numbers of positive and negative
fluctuations but faint sources are more numerous than the bright ones. Thus the number of faint sources
exceeding the threshold because they are on top of positive fluctuations is larger than that of brighter sources
falling below the threshold because they happen to be on top of negative fluctuations. This effect is known as
the Eddington bias.

The normalized distribution of true fluxes, $S$, of extragalactic sources, is usually well described by a power
law
\begin{equation} \label{eq:powlaw}
P(S|q) = k S^{-(1+q)}, \ \ \ S \geq S_m .
\end{equation}
If the slope $q$ is known, the maximum likelihood estimator of the true fluxes of the sources is easily
calculated \citep{hog&tur98} from the observed ones. Unfortunately, in many cases and in particular in the case
of the WMAP survey, this condition is not satisfied. For this reason, \citet{herr06} and LC07 have gone beyond
the results of the previous work by solving simultaneously for $q$ and $S$. We remind here only the asymptotic
limits of the estimators, valid in the high signal--to--noise regime, referring the reader to LC07 for a
detailed description of the calculations:
\begin{eqnarray} \label{eq:asym}
q & \simeq & \left[ \frac{1}{N}\sum_{i=1}^{N} \mathrm{ln} \left(
\frac{S^o_i}{S^o_m} \right) \right]^{-1} \\
S_i & \simeq & S^o_i \left[ 1- \frac{1+q}{r_i^2} \right],
\end{eqnarray}
\noindent where $S^o_m$ is the minimum observed flux and $r_i=S^o_i/\sigma_i$ is the signal to noise ratio of
the source. Firstly, we estimate the slope, $q$, from our observed fluxes [eq.~(\ref{eq:asym})] and then we use
it to calculate the correction for each source [eq.~(3)].

\subsection{Ground based measurements}\label{sec:sou_cal}

Many WMAP sources are calibrators  for the Australia Telescope Compact Array (ATCA) and/or for the VLA and, as
such, are continuously monitored at several frequencies. A list of 599 sources brighter than 500 mJy at 12 mm
(25 GHz), some of which have also data at 7 and 3 mm (43 and 100 GHz), is available on the ATCA
website\footnote{http://www.narrabri.atnf.csiro.au/calibrators/}. We have cross-correlated this list with the
complete NEWPS$_{5\sigma}$ sub-sample flux with $S_{23\rm GHz}\ge 2\,$Jy. Within a search radius of
$0.348^\circ$ (the beam-size of the WMAP K-band) we have found 49 associations, 47 of which have flux
measurements also at 7 mm and 49 at 3 mm. The numbers of  NEWPS ${5\sigma}$ detections for these 49 sources are:
48, 48, 43, and 8 at 33, 41, 61, and 94 GHz, respectively. The median flux density ratio for pairs of WMAP
channels was obtained using the \citet{KM58} estimator, taking into account also the ${5\sigma}$ upper limits.
The comparison with calibrators was performed in the space of WMAP and calibrator frequencies, defining a
$5\times3$ matrix of flux density ratios $R(\nu_W,\nu_c)$, with $\nu_W=[23,33,41,61,94]\,$GHz and $\nu_c=[25,
43, 100]\,$GHz.

Since the number of 94 GHz detections is too small, the application of the Kaplan-Meier estimator to data at
this frequency yielded unreliable results. We therefore decided not to take into account the
$R(94\,\hbox{GHz},\nu_c)$ values and to replace them with extrapolations from the other elements of the matrix.
In practice, for each of the calibrators frequencies  we have linearly fitted the
$R([23,33,41,61]\,\hbox{GHz},\nu_c(i))$ data (where i=1,2,3) and estimated the $R(94\,\hbox{GHz},\nu_c(i))$
elements by a direct extrapolation of the fits. Next, we estimated the $5\times5$ matrix $R(\nu_W,\nu_W)$ by
interpolating (we have used a 2-dimensional spline interpolation) the estimated $R(\nu_W,\nu_c)$ matrix. The
calibration factors for each  WMAP channel were finally obtained as the diagonal elements of the
$R(\nu_W,\nu_W)$ matrix. The results and their errors are given in the last column of Table~\ref{tb:cal_fac}.
Due to the uncertainty of the correction factor for the W band, hereinafter we will use only the corrected
fluxes relative to the K, Ka, Q and V WMAP bands for our statistical analysis.

\subsection{Self-calibration: methodology}\label{sec:auto_cal}

As stated in \S\,1, the uncertainty on flux calibration stems from our imperfect knowledge of the asymmetric
beam response functions. WMAP calibration is based on the brightness temperature of the CMB dipole; in order to
go from temperatures to fluxes it is necessary to multiply the temperature maps by a numerical factor that
depends on the frequency of observation (thus transforming the maps into units of Jy/sr, for example) and then
to multiply again by the effective area (solid angle) subtended by the point source. Therefore, a good knowledge
of the effective area of the beams is required. In LC07 the effective areas of the different WMAP beams were
calculated using the symmetrized beam profiles given by the WMAP
team\footnote{http://lambda.gsfc.nasa.gov/product/map/dr2/beam\_profiles\_ get.cfm}. Obviously, this is only a
first-order approximation. In particular, for the W and V bands the beams are known to be highly non-Gaussian
and non-symmetric. Thus, accurate flux determinations require the knowledge of {\it the real, non-symmetric
effective beam areas} at each source position. The WMAP team provided real beam maps, but it must be noted that
the \emph{effective} beam area at a given position in the sky is the result of a complicated composition of the
nominal beam over many pointings with different orientations at different times.

Fortunately, there is a simpler way to calculate the effective beam
area, only using WMAP data. In the absence of noise of any kind
(i.e. if only the point source signal is present), it would be
sufficient to use known point sources to measure directly the beam
area: the effective beam area would simply be the ratio of the total
observed source flux to its central value (i.e., the value
corresponding to the central pixel). The total NEWPS fluxes have been
calculated using the formula:
\begin{equation} \label{eq:total_flux}
F_i= A_b I_i(0),
\end{equation}
\noindent where $A_b$ is the effective beam area and $I_i(0)$ is the peak intensity for the $i$-th source,
directly calculated from the filtered image.

The SExtractor package \citep{sex} is a well-known software specifically designed for obtaining accurate
photometry of compact astronomical objects on images. This software allows us to estimate directly the total
flux of each source, $F_i$, without any assumption about the beam shape or size.

To minimize the errors on the estimated fluxes we need to select only {\it bright} sources, i.e. sources with a
high signal--to--noise ratio. Therefore, for each WMAP band we have selected the NEWPS$_{5\sigma}$ sources with
S$ \geq 1$ Jy and local S/N$\geq 5$ (in the real space) at that frequency; their numbers are 123, 75, 67, 28,
and 3 at 23, 33, 41, 61 and 94 GHz, respectively. Then, for each NEWPS source satisfying this selection
criterion we have projected a flat patch around its position. Subsequently, we have run SExtractor on the
resulting patches to obtain a direct estimation of the total flux and of its photometric error. Besides, for
each detected source the peak intensity and the background r.m.s. at its position has been obtained. Then, we
have calculated the effective area of the beam at the position of the source as $A_{b_i} = F_i/I_i(0)$. Finally,
we have computed the average of all the individual effective areas, weighted by the corresponding errors, for
each WMAP channel.

The results are presented in Table~\ref{tb:cal_fac}. In the first column we have the WMAP frequency band. The
second column shows the average effective beam areas calculated with the SExtractor method, and their errors,
while the third  column contains the corresponding beam area calculated using the symmetrized beam profiles
(this is the beam area used in NEWPS for unit conversion). If the estimated average effective areas apply to the
whole sky, the correction factor that should be applied to NEWPS fluxes is $A_{b,{\rm SEx}}/A_{b,{\rm sym}}$.
This factor is given in the fourth column of Table~\ref{tb:cal_fac}.  A comparison with the correction factors
estimated from calibrators (last column) shows a very encouraging agreement. Since for the W band we have only 3
sources that satisfy the selection criterion, the correction factor is much more uncertain than suggested by the
formal error, that only includes the dispersion of the estimated values.

For the following analysis we will {\it correct the NEWPS fluxes in LC07 multiplying them by the
self-calibration factors} listed in Table 1, except for the W band, that will not be considered further because
of the lack of a reliable calibration factor, the relatively high number of upper limits and its uncertain
completeness level.

\begin{figure}
\begin{center}
\includegraphics[width=0.35\textwidth,angle=90]{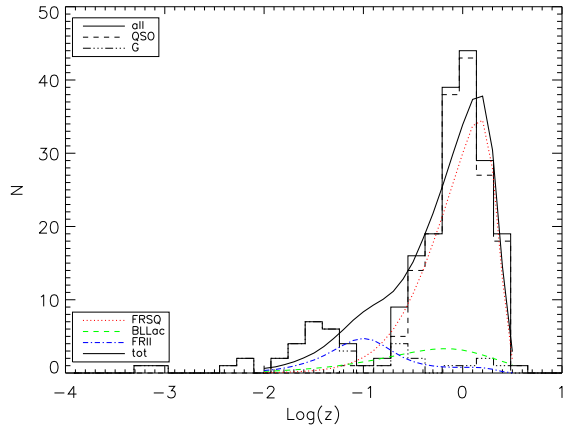}
\end{center}
\caption{Redshift distributions of the full $|b|>10^\circ$ sample, of QSOs and of galaxies (solid, dashed, and
dot-dashed histograms, respectively). The dotted, dashed, and dot--dashed curves display, for comparison, the
predictions of the model by \citet{dz05} for Flat-Spectrum Radio QSOs, BL Lacs, Fanaroff-Riley type II sources,
while the solid line shows the total.} \label{fig: z_distrib}
\end{figure}

\section{Optical identifications and redshift distribution}\label{sec:red_distrib_llf}

We have searched the NASA Extragalactic Database (NED\footnote{http://nedwww.ipac.caltech.edu/}) for optical
identifications of the 252 sources at $|b|> 5^\circ$ with $\ge 5\sigma$ detections at 23 GHz and flux density
above the completeness $S_{\rm 23 GHz}= 1.1\, \rm Jy$. The majority of sources (181, i.e. 72\%) are classified
as QSOs, 44 (17\%) as galaxies, 5 are Galactic objects (2 SNR, 2 HII regions, 1 PN), and the remaining ones (22,
i.e. 9\%) are unclassified. The galactic objects were obviously dropped from the sample.

The NED database also provided the redshifts of 215 sources (87\% of the extragalactic sample; 95\% (42) of
galaxies and 96\% (173) of QSOs). Although the redshift completeness is already rather high, it can be increased
restricting the sample to $|b|> 10^\circ$. In fact, in the Galactic latitude range $5^\circ < |b|< 10^\circ$
there are 22 extragalactic objects, only 8 of which have redshift measurements, not surprisingly because of the
substantial Galactic extinction. Our extragalactic sample comprises 225 objects at $|b|> 10^\circ$ and 207
(92\%) of them have redshifts (167 QSOs and 40 galaxies). Optical ($b_J$) magnitudes of these sources were
recovered from the SuperCOSMOS Sky Survey archives\footnote{http://www-wfau.roe.ac.uk/sss/}. The median $b_J$ is
16.7 for galaxies and 18.2 for QSOs.

The redshift distribution for the latter subsample is shown in Fig.~\ref{fig: z_distrib}, where the dashed,
dot-dashed, and solid histograms refer to QSOs, galaxies, and to the total, respectively. The median redshift of
the sample is 0.86 (0.052 for Galaxies only, 0.994 for QSOs). We have also plotted, for comparison, the redshift
distributions predicted by the \citet{dz05} model for different source populations. The agreement is generally
good, except for the dip in the data around $\log(z)=-1$, where the model predicts a little bump due to FRII
sources. The reason of this discrepancy is unclear. The most obvious option is that FRII sources are not
correctly modeled, but other possibilities, such as a large scale inhomogeneity in the distribution of bright
radio sources, cannot be ruled out.

\begin{figure}
\begin{center}
\includegraphics[width=0.45\textwidth]{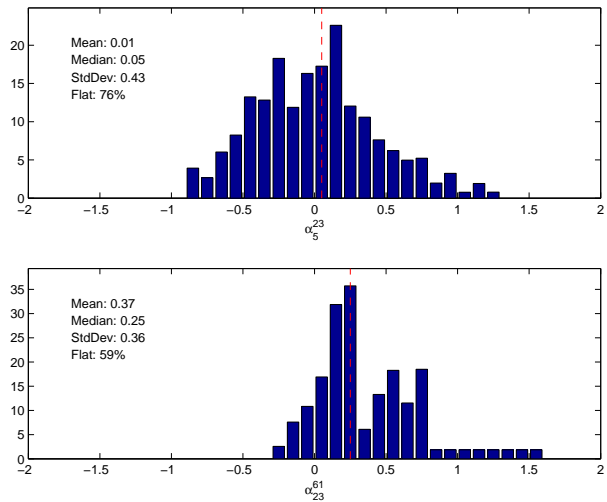}
\caption{Spectral index distributions for two different frequency
ranges: 5--23 GHz (upper panel) and 23--61 GHz (lower panel). For
both distributions the mean, median, standard deviation and
fraction of flat spectrum sources (see
\S~\ref{sec:sou_class}) are shown. The vertical red dashed lines correspond to the median values of the distributions.
See Section 4 for more details on the calculations.}%
\label{fig:alpha_dist}
\end{center}
\end{figure}

\section{Spectral index distributions and spectral classes }\label{sec:pop_spec}

To study the spectral properties of NEWPS sources we defined a complete sub-sample comprising all the 191 NEWPS
sources with $S_{23}\geq 1.3$ Jy and $S/N\geq 5$, with counterparts in the 5 GHz surveys (GB6 or PMN). We have
confined ourselves to fluxes somewhat above the 23 GHz completeness limit ($\sim 1.1\,$Jy) in order to have more
than 50\% $5\sigma$ detections at 61 GHz (we have a detection rate of 52\%); otherwise the uncertainties on the
spectral index distribution obtained from Survival Analysis become unacceptably large.

In addition to fluxes in the WMAP channels, we have included in our analysis 5 GHz measurements. Since the
angular resolution of the 5 GHz catalogues is much higher than that of any of the WMAP channels, we have
degraded the 5 GHz catalogue to the WMAP resolution. In practice, whenever more than one 5 GHz source are
present within a WMAP resolution element, we have summed up their 5 GHz fluxes, weighted with the WMAP beam
centered at the position of the NEWPS source. To correct for the contribution of the background of faint 5 GHz
sources we have selected a set of 4141 fields (control fields) at 23 GHz with area equal to that of a WMAP
resolution element at the considered frequency and devoid of NEWPS sources. The mean flux (120 mJy) of 5 GHz sources found
in control fields, weighted with the WMAP beam with axis towards the field center, has been
subtracted from the sum of 5 GHz fluxes (or the the flux of the only source) in the fields of NEWPS sources.

\subsection{Spectral index distributions}\label{sec:spec_distrib}

The spectral index distributions for the frequency intervals 5--23 GHz and 23--61 GHz are shown in
Fig.~\ref{fig:alpha_dist}. The latter distribution has been obtained by the Kaplan-Meier (1958) estimator,
taking into account also the upper limits on the fluxes, yielding lower limits on spectral indices. Note that,
after application of the Kaplan-Meier estimator it is no longer possible to distinguish between data and limits.
Interestingly, the distribution of $\alpha_{23}^{61}$ shows a hint of a second peak in the range 0.4--0.8, not
seen in the distribution of $\alpha_{5}^{23}$, consistent with the transition of a subset of sources from flat
to steep spectra. We have applied the non-parametric two sample tests implemented in the public ASURV
code\footnote{http://www.astrostatistics.psu.edu/statcodes/sc\_censor.html} to see whether the two distributions
are consistent with being drawn from the same parent distribution. This hypothesis is rejected with a high
significance level (more than $\sim 8\sigma$). As noted below, however, the difference may be due, to a large
extent, to a selection effect.

\begin{figure}
\begin{center}
\includegraphics[width=0.35\textwidth,angle=90]{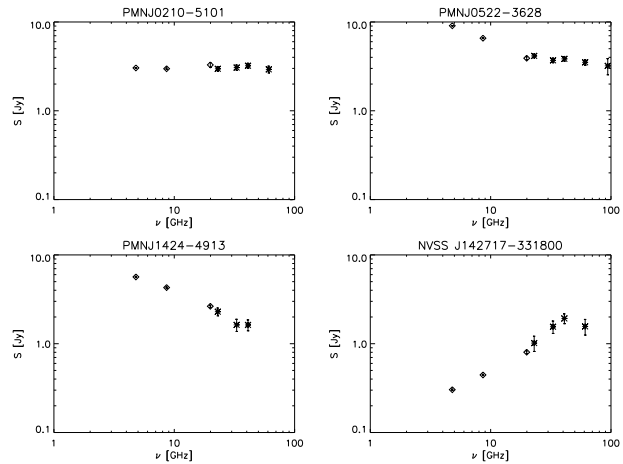}
\caption{Examples of typical radiosource spectra at mm wavelengths: a flat-spectrum source (\emph{top left panel});
 a steep-spectrum source (\emph{bottom left panel}); a source whose spectrum flattens at $\nu\sim 10$ GHz
 (\emph{top right panel}); a High Frequency Peaker (HFP) source (\emph{bottom right panel}). Data from the NEWPS Catalogue
 (\emph{asterisks}) and from the AT20G Survey (courtesy of the AT20G group, Massardi et al. (in preparation))
 (\emph{diamonds}).}
\label{fig:spectra}
\end{center}
\end{figure}

\begin{table}
  \begin{center}
   \begin{tabular}{ccc}
   Class & 5--23\,GHz & 23--61\,GHz  \\ \hline\hline%
   Flat (\%)    &76$\pm$6 &  59$\pm$6 \\ %
   Steep (\%)   &17$\pm$1 &  41$\pm$4 \\ %
   Inverted (\%)&7$\pm$1  &  0  \\ \hline\hline%
  \end{tabular}
  \end{center}
   \caption{Distribution of the 191 sources in our sample in three spectral classes for the 5--23 GHz and 23--61 GHz frequency intervals:
   flat--spectrum ($-0.5\leq\alpha\leq 0.5$),
   steep--spectrum ($\alpha>0.5$) and inverted--spectrum ($\alpha<-0.5$). See \S~\ref{sec:sou_class} for more details.} \label{tb:sp_class}
\end{table}

\subsection{Source spectral classes}\label{sec:sou_class}

Radio sources show a variety of spectral behaviours in this frequency range \citep[cf.][]{sadler06}.
Figure~\ref{fig:spectra} shows some examples:  a flat--spectrum (top left panel), a steep--spectrum (bottom left
panel), a spectrum flattening at $\sim 10$ GHz (top right panel) and a spectrum peaking at high radio frequency
(bottom right panel). The data are a combination of NEWPS fluxes with the AT20G measurements \citep{mass07}.

We have classified our sources into three main spectral classes: flat--spectrum sources
($-0.5\leq\alpha\leq0.5$), steep--spectrum sources ($\alpha>0.5$) and inverted--spectrum sources
($\alpha<-0.5$). Between 5 and 23 GHz flat--spectrum sources dominate the sample: they amount to 76$\pm 7$\%
(see Table~\ref{tb:sp_class}), in agreement with conclusions of previous works \citep{ben03b,ricci04,sadler06}.
However, at higher frequencies, we find a trend towards a steepening of the spectral indices. The 5-23 GHz
spectral index distribution has a median (mean) value $\alpha=0.05\pm0.04~(0.01\pm0.03)$, with a dispersion
$\sigma=0.43$, in full agreement with the results by \citet{ricci04} who used the multifrequency ATCA survey
data. The median (mean) spectral index between 23 and 61 GHz for the same sample, selected at 23 GHz, is
$\alpha=0.25\pm0.05~(0.37\pm0.04)$ (see Fig. \ref{fig:alpha_dist}); correspondingly, the fraction of the
steep--spectrum sources increases from $17\pm 1$\% to $41\pm 4$\%.

The quoted errors are Poissonian in the case of $\alpha_{5}^{23}$. As for $\alpha_{23}^{61}$, we must take into
account that its distribution includes a substantial fraction of lower limits, redistributed with the
Kaplan-Meier (1958) method. Clearly, limits cannot be given the same weight as actual measurements in the error
estimate. As discussed by Cantor (2001), in this case a simple and conservative estimate of the standard error
on the redistribution of limits is provided by the Peto formula
\begin{equation}\label{peto}
\sigma_{23}^{61} = P\left({1-P\over N_u}\right)^{1/2},
\end{equation}
where $P$ is the probability that $\alpha_{23}^{61}$ is larger than the mean (or median) value and $N_u= 91$ is
the number of lower limits. The probability $P$ is approximated by the fraction of sources with
$\alpha_{23}^{61}$ larger than the mean or the median ($P = 0.44 $ or $P = 0.50$, respectively). The global
error is computed as the quadratic sum of the Poisson error for the total number of data points (measurements
plus limits) plus the contribution given by eq.~(\ref{peto}). In the same way, we have estimated the errors on
the fraction of sources of steep- and flat-spectrum sources by identifying the probability $P$ with the fraction
of sources with spectral index above the boundary for their spectral class.

The 23 GHz selection obviously biases the spectral index distribution in favour of flat-/inverted values, which
yield brighter fluxes at higher frequencies. In the case of a Gaussian distribution, a dispersion
$\sigma_\alpha$ translates into an apparent flattening of spectral indices by
\citep{1984A&A...131L...1D,1984ApJ...287..461C}:
\begin{equation}
\Delta \alpha \simeq (1-\gamma)\sigma_\alpha^2\ln(23/5),
\end{equation}
$\gamma$ being the slope of differential counts. For an Euclidean slope ($\gamma=2.5$) and $\sigma_\alpha=0.3$
\citep[note that the distribution in Fig.~\ref{fig:alpha_dist} is broadened by errors on flux
measurements]{ricci04}, $\Delta \alpha \simeq -0.2$. Thus the observed steepening above 23 GHz may be largely a
selection effect.

On the other hand, a real steepening is expected in this frequency range since the absorption optical depth
decreases with increasing frequency at least as $\nu^{-2}$ so that an increasing fraction of the emitting
regions becomes optically thin, i.e. take on spectral indices $\alpha \simeq 0.7$--0.8 or steeper, because of
electron ageing effects.

\begin{figure}
\begin{center}
\includegraphics[width=0.45\textwidth]{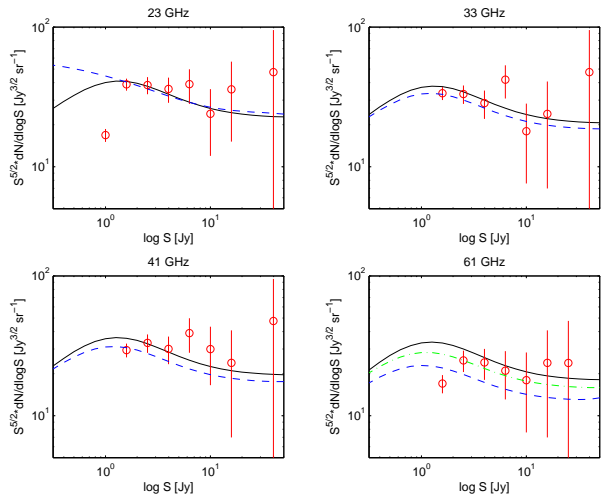}
\caption{Differential number counts, in bins of $\Delta\log(S)=0.2$, multiplied by $S^{\frac{5}{2}}$, of the
NEWPS$_{5\sigma}$ sources (circles with Poisson error bars). The solid curve shows, for comparison, the total
counts predicted by the \citet{dz05} model. The dashed lines correspond to the counts extrapolated from 5 GHz,
for the 23 GHz case, and from 23 GHz for the other frequencies, using the spectral index distributions of
\S\,\protect\ref{sec:spec_distrib}. The extrapolated counts are too low at 61 GHz, suggesting that the true
steepening of spectral indices above 23 GHz is less pronounced than implied by the distribution yielded by the
Kaplan-Meier estimator. A good fit is obtained damping down the secondary peak of the distribution of
$\alpha_{23}^{61}$ (Fig.~\protect\ref{fig:alpha_dist}) by a factor of 0.2 (dot-dashed line). }
\label{fig:counts_1}
\end{center}
\end{figure}

\begin{table}
  \centering
  \begin{tabular}{ccccc}
  \textbf{Freq}&\textbf{$S_{\rm lim}$}& \textbf{A ($\sigma_A$)}& \textbf{$\gamma ~(\sigma_{\gamma})$} & \textbf{Prob}\\
  (GHz)&(Jy)&($\hbox{Jy}^{-1}\,\hbox{sr}^{-1}$)&\\
  \hline\hline
    \textbf{23} & 1.1  & 34.4 (3.5)  & 2.36 (0.10)  & 0.84 \\
    \textbf{33} & 1.1  & 27.8 (3.2)  & 2.25 (0.11)  & 0.08 \\
                & 1.4  & 34.6 (5.0)  & 2.40 (0.13)  & 0.30 \\
    \textbf{41} & 1.1  & 30.9 (3.4)  & 2.41 (0.11)  & 0.99   \\
    \textbf{61} & 1.1  & 18.5 (2.8)  & 2.24 (0.14)  & 0.07   \\
                & 1.4  & 25.8 (4.5)  & 2.47 (0.15)  & 0.94 \\
    \hline\hline
 \end{tabular}
\caption{Parameters of the power-law approximation to the differential source number counts calculated from our
sample. The columns contain: 1. the frequency, 2. the lower flux limits, 3. the normalization, $A$, and 4. the
slope, $\gamma$, of the counts with their errors, and 5. the probability of exceeding the corresponding $\chi^2$
value if the model is correct. Only the data points on source counts plotted in Figure 5 and above the lower
flux limit, $S_{lim}$, of column 2 are used in this calculation.}\label{tab:counts2}
\end{table}

\section{Number counts}\label{sec:counts}

In Fig.~\ref{fig:counts_1} we have compared the differential source counts of NEWPS$_{5\sigma}$ sources at 23,
33, 41, and 61 GHz, normalized to the Euclidean slope and in bins of $\Delta\log(S)=0.2$, with the predictions
of the cosmological evolution model by \citet{dz05}. The agreement is generally good above the completeness
limits of the sample. However the flux density range probed with sufficient statistics is quite limited.

Over this range, the differential counts can be described by the usual power-law approximation:
\begin{equation}
\frac {dn}{dS}=A S^{-\gamma}.
\end{equation}
The best-fit values of the normalization, $A$, and of the slope, $\gamma$, are given, with their errors, in
Table~\ref{tab:counts2}, where the goodness of the fit is quantified by  the probability of exceeding the
corresponding $\chi^2$ value assuming that the model is correct. The normalization, $A$, is almost constant up
to 41 GHz, in keeping with the fact that the sample is dominated by sources with mean spectral index close to 0,
but slightly decreases at 61 GHz, as the effect of the high-frequency steepening.

Above 30 GHz, the values of $A$ are significantly higher than the estimates by \citet{waldram07}, based on
extrapolations of the 9C counts at 15 GHz. As discussed below, the discrepancy arises because the spectral index
distribution of the relatively faint sources of their sample is not appropriate at the high flux densities of
interest here.

\subsection{Number counts estimates by extrapolation from lower frequencies}\label{sec:extrapol_counts}

The spectral index distributions of \S\,\protect\ref{sec:spec_distrib}, combined with the observational
determinations of 5 GHz counts, extending over a very broad flux density interval, can be exploited to get model
independent estimates of the high frequency counts. Alternatively, the comparison of the extrapolated counts
with the observed ones provides a test of the Kaplan-Meier estimates of the spectral index distributions.

Under the assumption that the spectral index distribution is independent of flux density (only valid for limited
flux density intervals, see below), the source number counts can be extrapolated from the frequency $\nu_1$ to
the frequency $\nu_2$, convolving the counts at frequency $\nu_1$ with the probability that a source with
$S_1(\nu_1)$ has $S_2(\nu_2)$:
\begin{eqnarray} \label{eq:extrapol}
\frac{dN_2(S_2(\nu_2))}{d\log S_2(\nu_2)}d\log S_2(\nu_2)  = \nonumber  \\
=\int\frac{dN_1(S_1(\nu_1))}{d\log S_1(\nu_1)}\,P[S_1(\nu_1)|S_2(\nu_2)]\,d\log S_1(\nu_1).
\end{eqnarray}
The probability distribution $P(S_1(\nu_1)|S_2(\nu_2))$ is straightforwardly obtained from the corresponding
spectral index distribution.

However, to extrapolate the 5 GHz counts we need such distribution for sources selected at the same frequency,
while that of \S\,\protect\ref{sec:spec_distrib} refers to the 23 GHz selection. In practice, to derive the 23
GHz counts knowing those at 5 GHz and our distribution of $\alpha_{5}^{23}$ we need to invert
eq.~(\ref{eq:extrapol}) with $\nu_1$= 23 GHz and $\nu_2$= 5 GHz. This is a classical inversion problem (Fredholm
equation of the first kind) and we have used the Lucy's iterative method \citep{lucy} to solve it:
\begin{eqnarray} \label{eq:lucy}
\frac{dN(S_{23}(j))^{r+1}}{dN(S_{23}(j))^{r}} = \sum_{i}{\frac{dN^0(S_{5}(i))}{dN^r(S_{5}(i))}P(S_{23}(j)|S_{5}(i))},
\end{eqnarray}
\noindent where for the 5 GHz counts, $dN^0(S_{5}(i))$, we have used the description provided by the
\citet{dz05} model, that accurately reproduces the data. We have initialized the method (i.e for r=0) adopting
Euclidean counts, $dN/dS\propto S^{-5/2}$, but we have also checked that our results are insensitive to the
function used to initialize the process. Additionally, we have improved the stability of the method by smoothing
the $dN(S_{23}(j))^{r+1}$ in each iteration.

The results are shown by the dashed line in the upper left-hand panel of Fig.~\ref{fig:counts_1}: the
extrapolated counts are in good agreement with the direct estimate and with the \citet{dz05} model for $S>1$ Jy,
but are well above the model prediction at fainter fluxes. Since the downturn of the counts below $\simeq 1\,$Jy
is born out by the 9C counts at 15 GHz \citep{waldram03} and by the ATCA pilot survey counts at 18 GHz
\citep{ricci04}, the discrepancy must be attributed to a failure of the basic assumption underlying the
extrapolation. This is confirmed by the direct evidence of a change  with decreasing flux density of the
proportion of flat- and steep-spectrum sources contributing to the corresponding 5 GHz counts. This illustrates
the risks of extrapolations of counts with incomplete information about the statistical properties of source
populations.

Additional evidence of substantial changes of the spectral index distribution with decreasing flux density at
$\sim 20\,$ GHz is provided by \citet{waldram07}, who carried out multi-frequency follow-up of a sample complete
to 25 mJy at 15 GHz of extragalactic sources from the 9C survey. The median spectral index between 15 and 43 GHz
was found to be 0.89 for these sources, much fainter than ours. They also reported a spectral steepening with
increasing frequency.

The observed distribution of $\alpha_{23}^{61}$ can be directly exploited to extrapolate the 23 GHz counts to
higher frequencies. Since the \citet{dz05} model provides a good description of the counts (as determined by the
9C and ATCA surveys) also below the limit of WMAP ones at this frequency, we have used it to carry out the
extrapolations. The dashed lines in Fig.~\ref{fig:counts_1} show that the extrapolated counts tend to be lower
than those directly observed, particularly at 61 GHz. This is unexpected, since the observed spectral index
distributions are, if anything, broadened by the contributions of measurements errors on fluxes and this leads
to a flattening of the effective spectral index and, correspondingly, to an overestimate of the high frequency
counts, contrary to what is found here.

The likely explanation is that the Kaplan-Meier estimator over-populates the steep portion of the distribution
of $\alpha_{23}^{61}$. The observed counts are recovered if we scale down by a factor of 0.2 the distribution at
$\alpha_{23}^{61}>0.5$ (dot-dashed line in Fig.~\ref{fig:counts_1}).

\section{Conclusions} \label{sec:Conclusions}

We have presented and discussed a variety of statistical properties that characterize the extragalactic sources
in the NEWPS Catalogue (LC07). The flux calibration has been investigated with 2 different methods and
correction factors have been derived, except for the highest frequency WMAP channel, for which the available
information is insufficient.

A search of the NED database has yielded optical identifications for 89\% of the 252 sources in the complete
NEWPS sample with  $S_{23\rm GHz}\geq 1.1\,$Jy; 5 sources turned out to be Galactic, and were dropped. At high
Galactic latitudes ($|b|>10^\circ$) we obtain a sub-sample of 207 sources (92\%) with spectroscopic redshifts
(again from the NED), for a total number of extragalactic sources of 225 . The corresponding redshift
distribution is in generally good agreement with the predictions of the model by \citet{dz05}, except possibly
around $\log(z)\simeq -1$ where the model predicts a significant contribution from FR II sources, that is not
observed.

The distribution of spectral indices between 23 and 61 GHz, $\alpha_{23}^{61}$, was obtained with the
Kaplan-Meier estimator, taking into account the upper limits on 61 GHz fluxes. It is shifted towards steeper values
compared with the distribution of $\alpha_{5}^{23}$, obtained using the GB6 or PMN 5 GHz fluxes. Although the
steepening can be just a selection effect, hints of a real high-frequency steepening are noted.

The source number counts obtained from our sample have close to Euclidean slope and are in good agreement with
the prediction of the cosmological evolution model by \citet{dz05}.

We have shown that the use of our spectral index distribution to extrapolate the 5 GHz counts to high
frequencies leads to inconsistent results below $S_{23\rm GHz}\simeq 1\,$Jy because of a substantial change in
the mixture of flat- and steep-spectrum sources. The change, clearly visible comparing our distribution of
spectral indices with that obtained by \citet{waldram07} for a much fainter sample at the nearby frequency of 15
GHz, accounts for the difference between our high frequency counts and extrapolations from 9C counts. This is an
example of the risks inherent in extrapolations with incomplete information of source properties.

Extrapolations to higher frequencies of the 23 GHz counts using the estimated distribution of $\alpha_{23}^{61}$
tend to under-predict the observed counts, especially at 61 GHz, implying that the Kaplan-Meyer estimator
probably overpopulates the steep portion of the spectral index distribution. The observed counts are recovered
if the distribution is decreased by a factor of 0.2 for $\alpha_{23}^{61}>0.5$.

\section*{Acknowledgements}

We acknowledge partial financial support from the Spanish Ministry of Education (MEC) under project
ESP2004--07067--C03--01 and from the Italian ASI (contract Planck LFI Activity of Phase E2) and MUR. JGN
acknowledges a postdoctoral position at the SISSA-ISAS (Trieste). Thanks are due to the Australia Telescope 20
GHz group for the useful exchange of information and for the data used in Fig.~\ref{fig:spectra}. This research
has made use of the NASA/IPAC Extragalactic Database (NED) which is operated by the Jet Propulsion Laboratory,
California Institute of Technology, under contract with the National Aeronautics and Space Administration. It
has also made use of data obtained from the SuperCOSMOS Science Archive, prepared and hosted by the Wide Field
Astronomy Unit, Institute for Astronomy, University of Edinburgh, which is funded by the UK Particle Physics and
Astronomy Research Council.

\end{document}